# Intronic Alus Influence Alternative Splicing

Galit Lev-Maor[1], Oren Ram[1], Eddo Kim[1], Noa Sela[1], Amir Goren[1], Erez Y. Levanon[2], Gil Ast[1]*

1 Department of Human Molecular Genetics, Sackler Faculty of Medicine, Tel Aviv University, Tel Aviv, Israel, 2 Department of Genetics, Harvard Medical School, Boston, Massachusetts, United States of America

## Abstract

Examination of the human transcriptome reveals higher levels of RNA editing than in any other organism tested to date. This is indicative of extensive double-stranded RNA (dsRNA) formation within the human transcriptome. Most of the editing sites are located in the primate-specific retrotransposed element called Alu. A large fraction of Alus are found in intronic sequences, implying extensive Alu-Alu dsRNA formation in mRNA precursors. Yet, the effect of these intronic Alus on splicing of the flanking exons is largely unknown. Here, we show that more Alus flank alternatively spliced exons than constitutively spliced ones; this is especially notable for those exons that have changed their mode of splicing from constitutive to alternative during human evolution. This implies that Alu insertions may change the mode of splicing of the flanking exons. Indeed, we demonstrate experimentally that two Alu elements that were inserted into an intron in opposite orientation undergo base-pairing, as evident by RNA editing, and affect the splicing patterns of a downstream exon, shifting it from constitutive to alternative. Our results indicate the importance of intronic Alus in influencing the splicing of flanking exons, further emphasizing the role of Alus in shaping of the human transcriptome.

Citation: Lev-Maor G, Ram O, Kim E, Sela N, Goren A, et al. (2008) Intronic Alus Influence Alternative Splicing. PLoS Genet 4(9): e1000204. doi:10.1371/journal.pgen.1000204

Editor: Harmit S. Malik, Fred Hutchinson Cancer Research Center, United States of America

Received January 16, 2008; Accepted August 20, 2008; Published September 26, 2008

Copyright: © 2008 Lev-Maor et al. This is an open-access article distributed under the terms of the Creative Commons Attribution License, which permits unrestricted use, distribution, and reproduction in any medium, provided the original author and source are credited.

Funding: This work was supported by the Israel Science Foundation (1449/04 and 40/05), GIF, ICA (through the Ber-Lehmsdorf Memorial Fund), DIP, and EURASNET. EK is a fellow of the Clore Scholars Programme. AG is supported by the Adams Fellowship Program of the Israel Academy of Sciences and Humanities.

Competing Interests: The authors have declared that no competing interests exist.

* E-mail: gilast@post.tau.ac.il

9 These authors contributed equally to this work.

## Introduction

Alternative splicing enhances transcriptomic diversity and presumably leads to speciation and higher organism complexity, especially in mammals [1–3]. There are four major types of alternative splicing: exon skipping, which is the most prevalent form in higher vertebrates; alternative 5′ and 3′ splice site (5′ss and 3′ss) selection; and intron retention, which is the rarest form in both vertebrates and invertebrates [4,5]. At least 74% (and probably much more) of human genes that contain introns produce more than one type of mRNA transcript through alternative splicing; however, it is unclear which of these products are biologically functional and which are non-functional products of inaccurate splicing [3,6–8]. Thus, understanding the changes in the genome that dictate fixation of beneficial alternative splicing events or deleterious events (e.g., mutations leading to genetic disorders or cancer), or aberrant splicing events (noise in the system) is of great interest.

There are three known origins of alternatively spliced exons: 1) exon shuffling, which is a form of gene duplication [9–11]; 2) exonization of intronic sequences [12–16]; and 3) change in the mode of splicing from constitutive to alternative splicing during evolution [17,18]. One mechanism responsible for the shift from constitutive to alternative splicing is accumulation of mutations in the 5′ splice site region. Here we set out to examine additional mechanisms involved in the transition from constitutive to alternative splicing.

The primate-specific retrotransposons called Alu are ~280 nucleotides long. These are the most abundant retrotransposed elements in the human genome with about 1.1 million copies [16,19,20]. A large fraction of these Alu elements are located within intronic sequences, in both the sense and the antisense orientation relative to the mRNA, and can potentially form long regions of double-stranded RNA (dsRNA) [16,21–25]. There are indications that extensive secondary structure occurs between Alu elements. The evidence is embedded in analyses of the RNA editing mechanism: The human transcriptome undergoes extensive adenosine to inosine RNA editing [23,26,27]. RNA editing is directed by an adenosine deamination mechanism catalyzed by specific adenosine deaminases, termed dsRADs (double-stranded RNA adenosine deaminases) or ADARs [26,28,29]. ADARs are required for the formation of the dsRNA molecules that serve as substrates for the deamination process [30,31]. Hence, dsRNA regions formed between two Alus in opposite orientation within 2000 nucleotides of each other may serve as substrates for ADAR [21–25,27,28,32].

More than 90% of known editing sites are found in Alu elements and editing occurs in sense and antisense pairs of Alus but not in flanking non-Alu sequences [23,24,33]. Recently, it was shown that a pair of inverted Alus located within the 3′UTR of EGFP mRNA serves as a substrate for A-to-I RNA editing that stabilizes the binding of the p54 protein to the mRNA. This causes nuclear retention of the mRNA and the silencing of EGFP expression [34]. Another example comes from the NARF gene, where formation of Alu-Alu dsRNA and its subsequent editing generates a functional 3′ splice site that is essential for exonization of that intronic Alu; moreover, editing in that Alu eliminates a stop codon and modulates the strength of exonic splicing regulatory sequences

**Author Summary**

The human genome is crowded with over one million copies of primate-specific retrotransposed elements, termed *Alu*. A large fraction of *Alu* elements are located within intronic sequences. The human transcriptome undergoes extensive RNA editing (A-to-I), to higher levels than any other tested organism. RNA editing requires the formation of a double-stranded RNA structure in order to occur. Over 90% of the editing sites in the human transcriptome are found within *Alu* sequences. Thus, the high level of RNA editing is indicative of extensive secondary structure formation in mRNA precursors driven by intronic *Alu*-*Alu* base pairing. Splicing is a molecular mechanism in which introns are removed from an mRNA precursor and exons are ligated to form a mature mRNA. Here, we show that *Alu* insertions into introns can affect the splicing of the flanking exons. We experimentally demonstrate that two *Alu* elements that were inserted into the same intron in opposite orientation undergo base-pairing, and consequently shift the splicing pattern of the downstream exon from constitutive inclusion in all mature mRNA molecules to alternative skipping. This emphasizes the impact of *Alu* elements on the primate-specific transcriptome evolution, as such events can generate new isoforms that might acquire novel functions.

(ESRs). Interestingly, the nucleotides surrounding the editing site are important not only for editing of that particular site but also for editing at other sites located downstream in the same exon. It was also shown that the C nucleotide thought to pair with the edited site on the dsRNA is important for editing [35]. There is emerging evidence that the secondary structure of precursor mRNA plays a role in regulation of alternative splicing. However, in most studies the double-stranded structure was made up of only 10–40 base pairs and sequestered exonic or splice site sequences [36–47].

In this study, we have bioinformatically and experimentally evaluated the effects of intronic *Alu* elements on splicing. We found that different regulatory constraints act on *Alu* insertions into introns that flank constitutively or alternatively spliced exons. We further demonstrated that two *Alu* elements which were inserted into introns in opposite orientation have the potential to undergo base-pairing, as evident by RNA editing, and affect the splicing patterns of a downstream exon, by shifting it from constitutive to alternative. Finally, as *Alu* elements are abundant in introns, the findings we present suggest that the effect of intronic *Alu* elements on the transcriptome could be substantial, and could result in transcriptomic novelties. The new isoforms could then be subjected to purifying selection which will determine their fixation.

## Results

### Genome-Wide Analysis of *Alu*s within Intronic Sequences

To examine potential effects of insertion of *Alu* elements into introns on splicing of the flanking exons, we downloaded data of human introns (hg18) and *Alu* elements and determined the intersected set using the UCSC genome browser and GALAXY [48,49]. Overall, 730,622 *Alu* elements that reside within introns and 185,534 introns were extracted. This analysis showed that there are 85,126 introns that contain at least one *Alu* element; of these, 5009 introns contained at least two *Alu* elements in opposite orientation. The median length of introns containing at least one *Alu* element is 3829 base pairs (bp), whereas the median length of introns that do not contain an *Alu* is 521 bp (for intron length distribution see Figure S1). This suggests that double strand formation as a result of base pairing between two nearby *Alu* elements in opposite orientation might be common.

### Bioinformatic Analysis of *Alu*-Editing within Intronic Sequences

An antisense *Alu* and a sense *Alu* that are within 2000 nucleotides of each other can form dsRNA and be subjected to mRNA editing [21–24,50]. To examine the configurations of possible *Alu*-*Alu* dsRNA we extracted data on 10,113 nucleotides that undergo mRNA editing in the human genome from Levanon et al. [24]. Intersection with data on all *Alu* elements in the human genome and human RefSeq intronic sequences yielded 953 *Alu* elements that are embedded in intronic sequences that undergo mRNA editing (see Materials and Methods). For each of the 953 edited *Alu* elements, the nearest *Alu* element in the opposite orientation was identified. For the vast majority of edited *Alu* elements (880 out of the 953; 92%), the closest *Alu* element in the opposite orientation resided within the same intron, with an average distance of 682 bp from the edited *Alu*. However, we found 73 cases (8%) where the nearest *Alu* element in the opposite orientation was in a different intron; in 61 of these cases it is at least 500 bp closer to the edited *Alu* element than the nearest *Alu* element in the opposite orientation in the same intron (Figure S2). In these cases, the average distance between the edited *Alu* and the nearest *Alu* in the opposite orientation is 1993 bp. In fact, in 43 out of the 61 cases the distance was less than 2000 bp (averaging 1122 bp). This close proximity between the two *Alu* elements, along with the evidence that at least one of them undergoes editing, suggests that these regions may base pair.

Since in 92% of edited *Alu* elements, the closest *Alu* element in the opposite orientation resided within the same intron, we decided to examine the splicing process in these cases. But first, we set to examine the distribution of *Alu* elements within datasets of exons conserved within human and mouse having different splicing patterns.

### *Alu*s More Common in the Flanking Introns of Alternatively Spliced Exons that Changed Their Splicing Pattern after the Primate-Rodent Split

We analyzed three datasets of human-mouse orthologous exons and their flanking introns and exons: 1) conserved constitutively spliced exons (constitutively spliced in both species, 45,553 exons), 2) conserved alternatively spliced exons (alternatively spliced in both species, 596 exons), and 3) exons that are alternatively spliced in human and constitutively spliced in mouse (species-specific alternative exons; 354 exons). Analysis of *Alu* insertions into introns flanking these exons revealed that species-specific alternative exons exhibited the highest level of *Alu* insertions, followed by conserved alternative exons; the group with the fewest intronic insertions were conserved constitutively spliced exons. We calculated the density of *Alu* insertions, namely the number of *Alu*s divided by the total intron length (and then multiplied by 1000 for convenience), in order to control for the fact that different intronic lengths might influence *Alu* insertion (see Materials and Methods). On average, 0.42 *Alu* elements were found per 1000 bp within the upstream introns and 0.41 *Alu* elements were found per 1000 bp within the downstream introns of constitutively spliced exons; 0.49 and 0.44 *Alu*s per 1000 bp were found in the upstream and in the downstream introns of alternatively spliced exons, respectively; and 0.66 and 0.65 *Alu*s were found per 1000 bp in the upstream and in the downstream introns of species-specific alternatively spliced exons, respectively. Thus, the number of *Alu* elements present in species-specific alternative exons differed significantly from that found in constitutively spliced exons (p-value = 7.16E-10, p-value = 5.22E-09, for upstream and downstream introns, respectively) and also differed from that in the alternatively spliced exons (p-value = 0.000503, p-value = 0.000014, for upstream and downstream introns, respectively).

Furthermore, analysis of the distribution of antisense and sense Alus upstream of conserved constitutively spliced exons revealed a selection against the presence of Alu elements adjacent to exons, specifically, against Alu elements in the antisense orientation (Figure 1). There are significantly fewer antisense Alus compared to sense Alus within 150 bp of constitutively spliced exons (p-value = 0.000012). Examination of the downstream intron did not reveal a significant bias (p-value = 0.056). This implies that a selective pressure exists against insertion of Alus in close proximity upstream to constitutively spliced exons; this bias is stronger against Alus in the antisense orientation than against the sense orientation.

B1 is a rodent-specific retrotransposed element of ~150 nucleotides that has the same ancestral origin as Alu: the 7SLRNA [51,52]. Like Alus, large numbers of B1 elements (a total of 331,015) reside within intronic sequences [16]. We found 236,036 B1 elements within 177,766 mouse introns that are found in GenBank (see Materials and Methods). In mouse, 70,516 introns (39.6%) contain B1 elements. Overall, there are 1.32 B1 elements per intron. The median length of introns containing at least one B1 element is 3278 bp, whereas the median length of introns that do not contain B1 is 636 bp. These results indicate that Alu and B1 containing introns are substantially longer than other introns.

We observed a significant correlation between the number of insertions of Alu and B1 elements within orthologous introns (Pearson correlation coefficient of 0.73 with p-value <0.0001). Namely, orthologous introns show the same tendencies for Alu and B1 insertion, although these events happened independently after the split of the mouse and human lineages. We then set out to analyze whether insertion of B1 into rodent introns was biased in terms of location and orientation as was the case for Alu in primates. Analysis of B1 insertions within the flanking introns of conserved constitutively spliced exons, conserved alternatively spliced, and species-specific alternatively spliced exons yielded the same trend as that of Alu insertions in human. There was no statistical difference in the density of B1 between conserved constitutively spliced and conserved alternatively spliced exons;

however, the upstream introns of the 258 species-specific exons (alternatively spliced in mouse, but constitutive in human) were significantly more enriched with B1 elements than were the upstream introns of conserved constitutively spliced exons (p-value = 0.0012) or constitutively spliced downstream introns (p-value = 0.014). This was also the case when the regions upstream of exons that are alternatively spliced in mouse and constitutively spliced in human were compared to the upstream introns of conserved alternatively spliced exons (p-value = 0.042) but not the downstream introns (p-value = 0.155). Therefore, insertion of retrotransposed elements into intronic sequences is correlated with the mode of splicing of the flanking exons.

Five reports [53–57] indicate that de novo Alu insertions into intronic sequences in antisense orientation and in close proximity to the affected exon (between 19–50 nucleotides) cause the downstream exon to shift from constitutive splicing to full exon skipping (three cases) or to alternative splicing (two cases) (Table 1). This effect of Alu elements on adjacent exons may be due to the Alu structure. Alu elements are comprised of two very similar segments, termed left and right arms. When an Alu is located in a gene in the antisense orientation and transcribed it contributes two poly-T stretches to the mRNA precursor. These poly-T regions might act as polypyrimidine tract (PPT) and, in combination with downstream 3' and 5' pseudo splice sites, might act as pseudo-exon [58]. Hence, such antisense Alus that act as pseudo-exons might compete with nearby exons for the binding of splicing factors. These five cases of de novo Alu insertions imply that Alus located in close proximity to exons might affect splicing of adjacent exons. This and the finding of de-novo Alu insertions that affect splicing imply that this is an ongoing evolutionary process, which may result in novel transcripts that are deleterious and inflict genetic diseases. On the other hand, a shift in the splicing pattern from constitutive to alternative might be advantageous in some cases, and could enable testing new mRNA options without eliminating the old ones. Moreover, such a shift could introduce a premature termination codons enabling the expression of truncated proteins at certain needed times or in specific cell types and could be delicately regulated by the levels of splicing regulatory proteins [59,60].

In order to determine how many alternatively spliced exons are potentially regulated by the insertion of an antisense Alu, we used the alternative splicing track in the UCSC genome browser ([48]

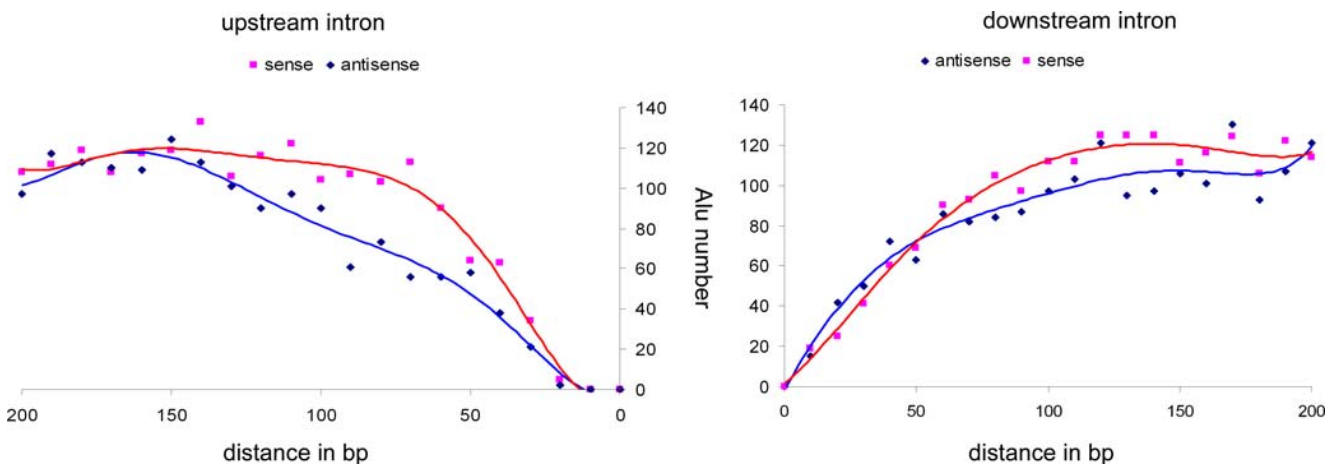

**Figure 1. Bioinformatic analysis of Alu insertion within the flanking introns.** Conserved constitutively spliced exons were analyzed for the differences in the location of antisense and sense Alus within the upstream and downstream introns (left and right panels, respectively). The x-axis is the distance in base pairs from the exon; the y-axis is the number of Alus found within this distance. Antisense Alus are marked in blue and sense Alus are marked in red.
doi:10.1371/journal.pgen.1000204.g001

Table 1. Diseases resulting from Alu insertion within an intron.

| Gene | Disease | Alu | Intron insertion | Orientation | Distance from SS | Effect | Reference |
|---|---|---|---|---|---|---|---|
| Fas | ALPS syndrome | Sb1 | 7 | antisense | 50 bp upstream of exon 8 | Skipping of exon 8 | Tighe et al., 2002 [55] |
| GK | Glycerol kinase deficiency | Y | 4 | antisense | 52 bp upstream of exon 5 | Alternative Splicing | Zhang et al., 2000 [57] |
| FGFR2 | Apert syndrome | Ya5 | 8 | antisense | 19 bp upstream of exon 9 | Alternative Splicing | Oldridge et al., 1999 [54] |
| NF1 | Neurofibromatosis type1 | Ya5 | 5 | antisense | 44 bp upstream of exon 6 | Skipping of exon 6 | Wallace et al., 1991 [56] |
| Factor VIII | Hemophilia A X-linked severe bleeding disorder | Yb9 | 18 | antisense | 19 bp upstream of exon 19 | Skipping of exon 19 | Ganguly et al., 2003 [53] |

doi:10.1371/journal.pgen.1000204.t001

see also Materials and Methods). In 269 events (~1.5% out of 17,151 alternatively spliced cassette-exons), an antisense Alu was found within 100 bp upstream of an exon (150 have additional sense Alu within 2000 bp), 491 (~2.8%) events in which an antisense Alu was found within 150 bp (273 have additional sense Alu within 2000 bp), and 689 (~4%) events in which an antisense Alu was found within 200 bp (373 have additional sense Alu within 2000 bp). Out of these 689 alternative exons, 525 (76.1%) are conserved between human and mouse (23 events were recorded as alternatively spliced also in mouse alternative splicing track in version mm9). Within the human genome, almost 85% of alternative cassette exon skipping events are conserved in mouse, however only 76% of the cassette exon skipping events that have an adjacent Alu in opposite orientation are conserved within mouse genome. This is statistically significant, implying that there is a bias for Alu in antisense orientation in the regulation of alternative exons within non-conserved alternative splicing events ($\chi^2$ test p-value = $1.6 \times 10^{-8}$).

### Intronic Alus Affect Splicing of Flanking Exons

The above results suggest that stable insertion of Alus into introns is associated with the mode of splicing of the flanking exons—especially the downstream exon—and that most Alu-Alu dsRNA is formed between sequences within the same intron. To test this hypothesis, exon 3 of the human RABL5 gene was analyzed experimentally to examine the connection between intronic Alu and alternative splicing. A minigene containing exons 2 through 4 of the human RABL5 gene (a conserved gene within all vertebrate genomes) was cloned. Exon 3 of RABL5 is alternatively spliced in human and constitutively spliced in mouse, rat, dog, chicken, and zebrafish (see Figure 2 in [17]). Based on the phylogenetic relationships among the analyzed organisms, we conclude that the alternatively spliced variant is a derived form and the constitutively spliced variant is the ancestral one.

Six Alus have been inserted into the flanking introns of exon 3 since the last common ancestor of human and mouse: two in the upstream intron and four in the downstream intron (Figure 2A). Alu4, which is located in the downstream intron, resulted from insertion of an Alu within another Alu and will be regarded as one Alu (see also Text S1). The minigene was transfected into 293T cells and the splicing products were examined following RNA extraction and RT-PCR analysis. Exon 3 in the RABL5 minigene is alternatively spliced with approximately 40% inclusion (Figure 2B, lane 1). Removal of all intronic Alus shifted splicing from alternative to constitutive (Figure 2B, compare lanes 1 and 2). This indicates that the insertion of Alus into the flanking introns during primate evolution shifted exon 3 splicing from constitutive to alternative.

Our experiments revealed that the orientation and position of the Alus within the upstream intron affected splicing of exon 3. Deletion of the Alus in the upstream intron, namely Alu1 and Alu2 (Δ1+2), had the same effect as deleting all Alus (Figure 2B, lanes 7 and 2). The same effect was observed if one of the Alus was deleted and the other was replaced with a 270-nucleotide non-Alu intronic sequence (Figure 2B, lanes 18 and 21, see also Text S2). The replacement of each of the intronic Alus with a non-Alu intronic sequence of a similar length eliminated the possibility that the effect observed after deletion related to shortening of the intron.

The Alus in the downstream intron, however, had a little or no effect on splicing (Figure 2B, lane 13). Taken together, it seems that the shift from constitutive to alternative splicing in the lineage leading to human is mediated mainly by Alus 1 and 2. Deletion of Alu1 or replacement with a 270-nucleotide non-Alu intronic sequence resulted in almost complete exon skipping (Figure 2B, lanes 3 and 17, respectively). Replacement or deletion of Alu2 resulted in constitutive exon splicing (lanes 4 and 20).

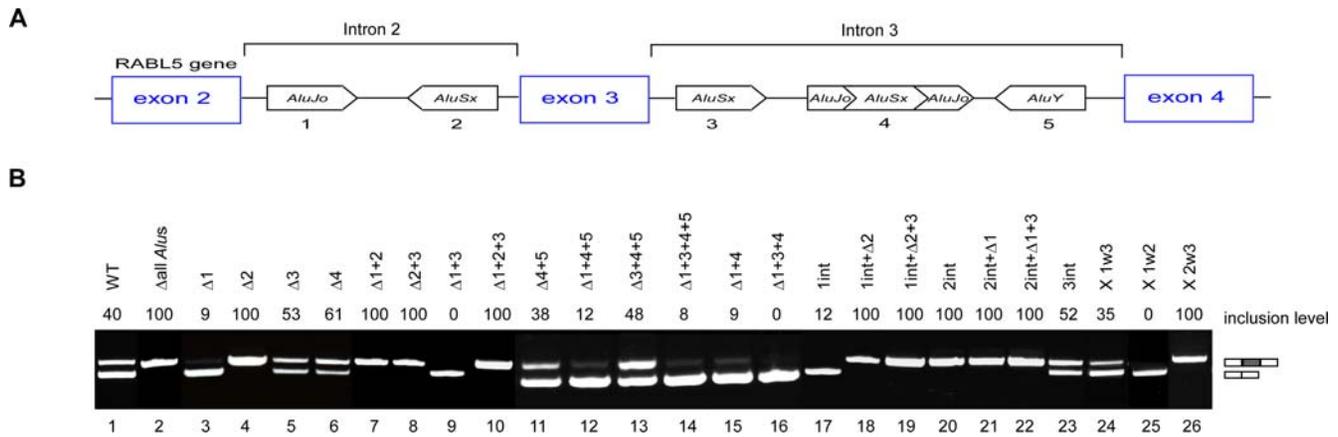

**Figure 2. The effect of intronic Alus on the splicing of a flanking exon.** (A) Schematic illustration of the RABL5 minigene containing three exons and two introns. The intronic Alus (1 through 5) are marked by boxes with a point indicating the orientation of the Alu relative to the pre-mRNA. Alu4 is an AluSx inserted between the two arms of AluJo. (B) The indicated wild-type (wt) and mutant plasmids were transfected into 293T cells, total RNA was extracted, and splicing products were separated on a 2% agarose gel after RT-PCR analysis. Lane 1, splicing products of wt RABL5 minigene. Lanes 2–26, splicing products of the indicated mutants. The following abbreviations were used: Δ indicates deletion of the specified Alu element, X 1w3 indicates replacement of Alu1 with Alu3 (i.e., the sequence of Alu3 was inserted instead of that of Alu1 and in the same orientation as Alu3), and 1int specifies replacement of the Alu1 sequence with a non-Alu intronic fragment. The two mRNA isoforms are shown on the right. Numbers on top of the gel indicate percentage of exon inclusion as determined using ImageJ software. PCR products were sequenced.
doi:10.1371/journal.pgen.1000204.g002

Interestingly, Alu1 and Alu2 have opposite effects on splicing. Deletion of both Alus has the same effect as deleting only Alu2. Therefore, we concluded that Alu2 is dominant over Alu1. The dominance of Alu2 is also supported by two other observations. First, if all Alus except Alu2 are removed, we observe almost total exon skipping (Figure 2B, lane 14). This indicates that Alu2 is a negative regulator of exon 3 recognition, unless Alu1, which is in opposite orientation to Alu2, is present (compare lanes 13 and 14). Furthermore, deletion or replacement with a 270-nucleotide non-Alu intronic sequence of Alu2 in combination with any additional intronic Alus leads to constitutive splicing (lanes 7, 8, 10, 18–22). In the absence of Alu1 and the presence of Alu2, the dominance of Alu2 over the other Alus is observed, leading to exon skipping (lanes 9, 12, 14, 15, and 16). As expected, in the presence of both Alu1 and Alu2, deletion of Alus from the downstream intron had a marginal effect on splicing (Figure 2B, lanes 5, 6, 11, and 23).

We demonstrated that the antisense orientation of Alu2 is essential for alternative splicing of exon 3. We first noted that the exact Alu family is not an important factor in determining splicing pattern: replacement of Alu1 of the Jo family with the sequence of Alu3 from the Sx family (both Alus are in the sense orientation) did not affect the splicing pattern (Figure 2B, lane 24). Thus, the important factor is the presence of Alu1 in sense orientation. Our analysis showed that only when Alu2 is in the antisense orientation and Alu1 is in the sense orientation is alternative splicing of exon 3 observed (Figure 2B, lanes 25–26). These results indicate that the two Alus in the upstream intron regulate alternative splicing of exon 3, whereas the three downstream intronic Alus have no apparent effect on splicing of that exon. Moreover, Alu2 in the antisense orientation suppressed inclusion of exon 3, whereas Alu1 in the sense orientation antagonized the effect of Alu2.

### Formation of Double-Stranded RNA between the Two Intronic Alus

How do the two intronic Alus regulate alternative splicing of exon 3? It is apparent that if Alu2 alone is present in the mRNA precursor, the exon is always skipped. We therefore postulated that in a population of the mRNA precursors that contain both Alu1 and Alu2, the two might form dsRNA formation (Figure 3A). This sequestration leads to exon inclusion; in the fraction of mRNA precursors with no base pairing between Alu1 and Alu2 the exon is skipped. To support this hypothesis, we examined whether RNA editing occurred in intron 2; editing would be indicative of formation of dsRNA. We searched the human EST/cDNA dataset and found five different mRNAs sequences containing intron 2 and comparison with genomic sequence indicated extensive RNA editing in both Alu1 and Alu2 (marked in red in Figure 3B). The region of the editing was found to be in the middle of both Alus. This suggests that these two Alu regions are in a double-stranded form.

To further confirm the formation of dsRNA, we generated cDNA from a neuroblastoma cell-line using specific primers (Figure 3C). By using primers that hybridize in the exons flanking intron 2, we were able to observe a small population of mRNA molecules that contain intron 2; the majority of mRNAs are spliced (Figure 3C lane 1). We enriched the intron-containing fraction using primers that hybridize within the intron and within the downstream exon (Figure 3C, lane 2). Sequencing of the higher molecular weight PCR product using primer to Alu2 allowed us to identify four editing sites within it (Figure 3D).

To confirm the importance of pairing between Alu1 and Alu2 on editing, we used the ΔAlu1 mutant (see Figure 2 lane 3) that led to a full exon skipping and examined the effect on editing within Alu2. There is one editing site within Alu2 that is dependent on the presence of Alu1; without Alu1 no editing at this site was observed (Figure 3E, the site is also highlighted in green in Figure 3B). The other putative editing sites found in EST/cDNA show relatively low level of editing in the minigene.

### The Importance of the Distance between the Intronic Alus and Exon 3

We next set to examine if the distance between exon 3 and the intronic Alus and the distance between Alu1 and Alu2 were important factors in splicing of exon 3. Alu2 is located 24 nucleotides upstream of exon 3. We identified the putative branch

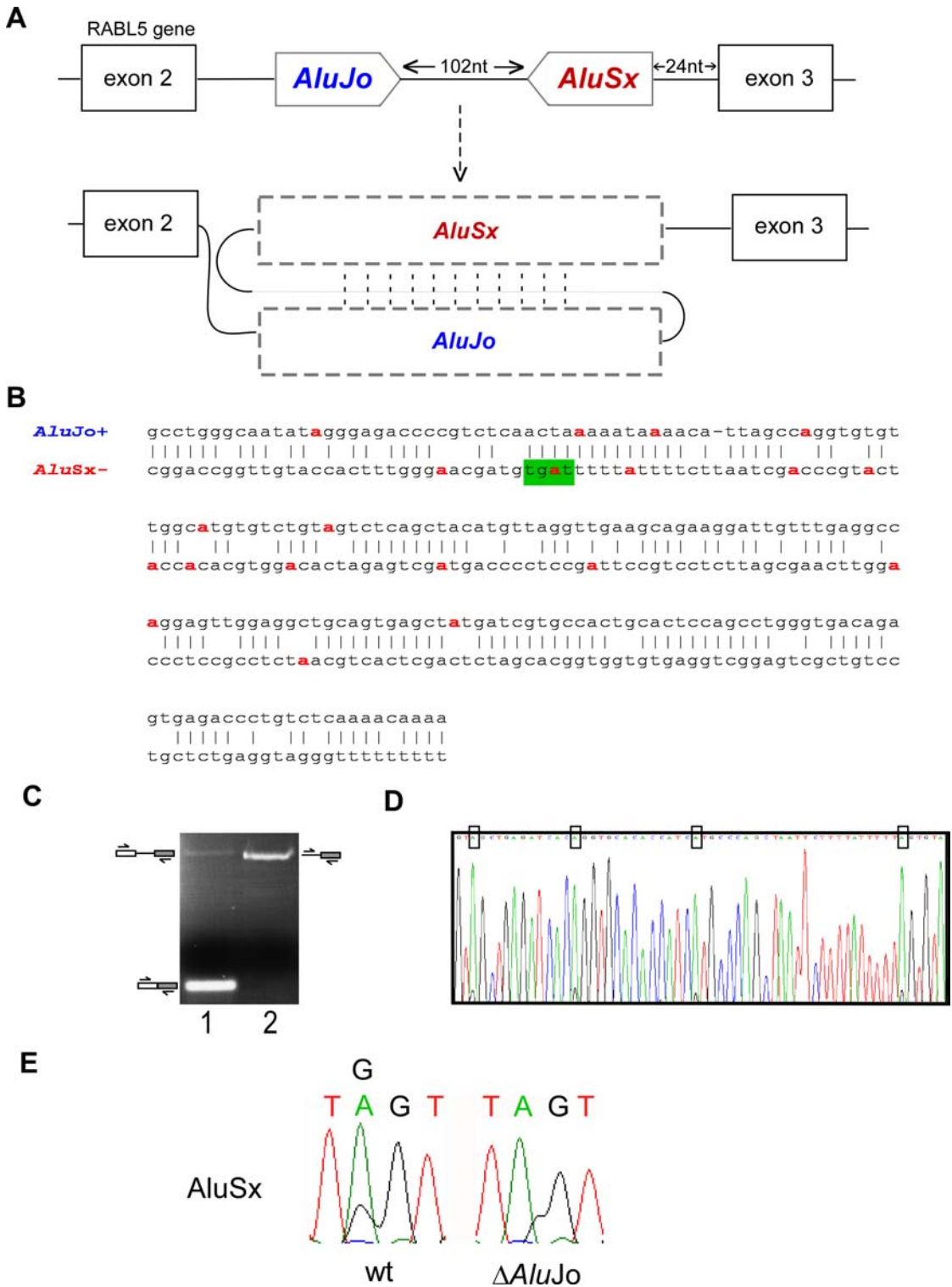

**Figure 3. Editing sites within the intronic Alus.** (A) Schematic illustration of exons 2 to 3 of the *RABL5* gene. Exons are depicted as black boxes; the intronic *Alus*, derived from *Alu*Jo and *Alu*Sx, in sense and antisense orientations, respectively, are shown in the middle gray-shaped boxes. The intronic, antisense *Alu* sequence (*Alu*Sx) is 102 nucleotides downstream of the sense *Alu*Jo and *Alu*Sx is 24 nucleotides upstream of the junction of exon 3. Sense and antisense *Alus* are expected to form a double-stranded secondary structure, thus allowing RNA editing. (B) Editing sites were inferred from alignment of five cDNAs (accession numbers BC050531, BC038668, BI547904, BI548328, and DB495755) to the human genomic DNA. RNA editing occurs at eight positions within the antisense *Alu*Sx and at eleven positions within the sense *Alu*Jo. Based on these editing sites, the pairing between the sense and antisense *Alu* sequences was inferred (upper and lower lines, respectively). The region in which editing occurs starts at the middle of the right arm (position 232 in the *Alu*Sx consensus) and ends at the beginning of the left arm of *Alu*2 (position 101 in the *Alu*Sx consensus). Panel B shows only this corresponding region, while the entire *Alu-Alu* potential dsRNA is shown in Figure S3. (C) To further confirm the editing activity, total RNA was extracted from a neuroblastoma (SH-SY5Y) cell-line and treated with DNaseI, followed by RT-PCR analysis using primers

to exon 2 and exon 3, and to intron 2 and exon 3 (lanes 1 and 2, respectively; see also Materials and Methods). The PCR products were sequenced. (D) The upper PCR product shown in panel C lane 2 was cloned and sequenced. The Chromas sequence is shown with the editing sites, found in AluSx, marked by boxes. (E) Editing in *Alu*2 requires the presence of *Alu*1. Wild type RABL5 minigene (WT) and a mutant in which *Alu*1 was deleted (ΔAluJo) were transfected into 293T cells. RNA was extracted, treated with DNase I, and amplified using set of primers flanking *Alu*2 and designed to amplify only exogenic transcripts. Sequencing chromatograms of four nucleotides in AluSx are shown (this editing site is also marked in green in panel B).
doi:10.1371/journal.pgen.1000204.g003

site of intron 2 and inserted an 800-nucleotide non-*Alu* intronic sequence upstream of the branch sequence and downstream of intronic *Alu*1 and *Alu*2 (marked B in Figure 4A; see Text S3). This insertion caused a shift from alternative to constitutive inclusion of exon 3 (Figure 4B, compare lane 1 and 2). Only when this insertion was shortened to less than 68 nucleotides did we begin to detect restoration of alternative splicing of exon 3; the level of skipping was further elevated when the inserted sequence was shortened to 56 or to 44 nucleotides (Figure 4B, lanes 3–9). To rule out the possibility that the sequence that was inserted contained intronic splicing regulatory sequences, we designed a fragment of 25 nucleotides free from known splicing regulatory sequences (see Materials and Methods). We inserted this sequence into site B and also duplicated and triplicated this sequence to generate 50 and 75 nucleotides insertions. The longer is the inserted sequence, the higher is the inclusion level (Figure 4B, lanes 10–13). This indicates that the distance between the intronic *Alu*2 from exon 3 affects the mode of splicing.

We also examined the effect of the distance between *Alu*1 and *Alu*2 on the splicing of exon 3. The insertion of the same fragment of 800 nucleotides between the two elements (marked A in Figure 4A) led to substantial reduction in the inclusion level of exon 3, although alternative splicing was still observed (Figure 4C, compare lane 1 and lane 2). We then shortened this sequence, ultimately to 24 nucleotides; when the distance was shorter than 550 nucleotides, almost complete inclusion of exon 3 was observed (Figure 4C lanes 2–9). Our results indicate that the distance between *Alu*1 and *Alu*2 is important for maintaining the alternative splicing of exon 3; however, it is not as important as the distance between the intronic *Alu* elements and exon 3. We also note that increasing the distance between exon 3 and *Alu*2 leads to exon inclusion, whereas increasing the distance between *Alu*1 and *Alu*2 enhances exon skipping.

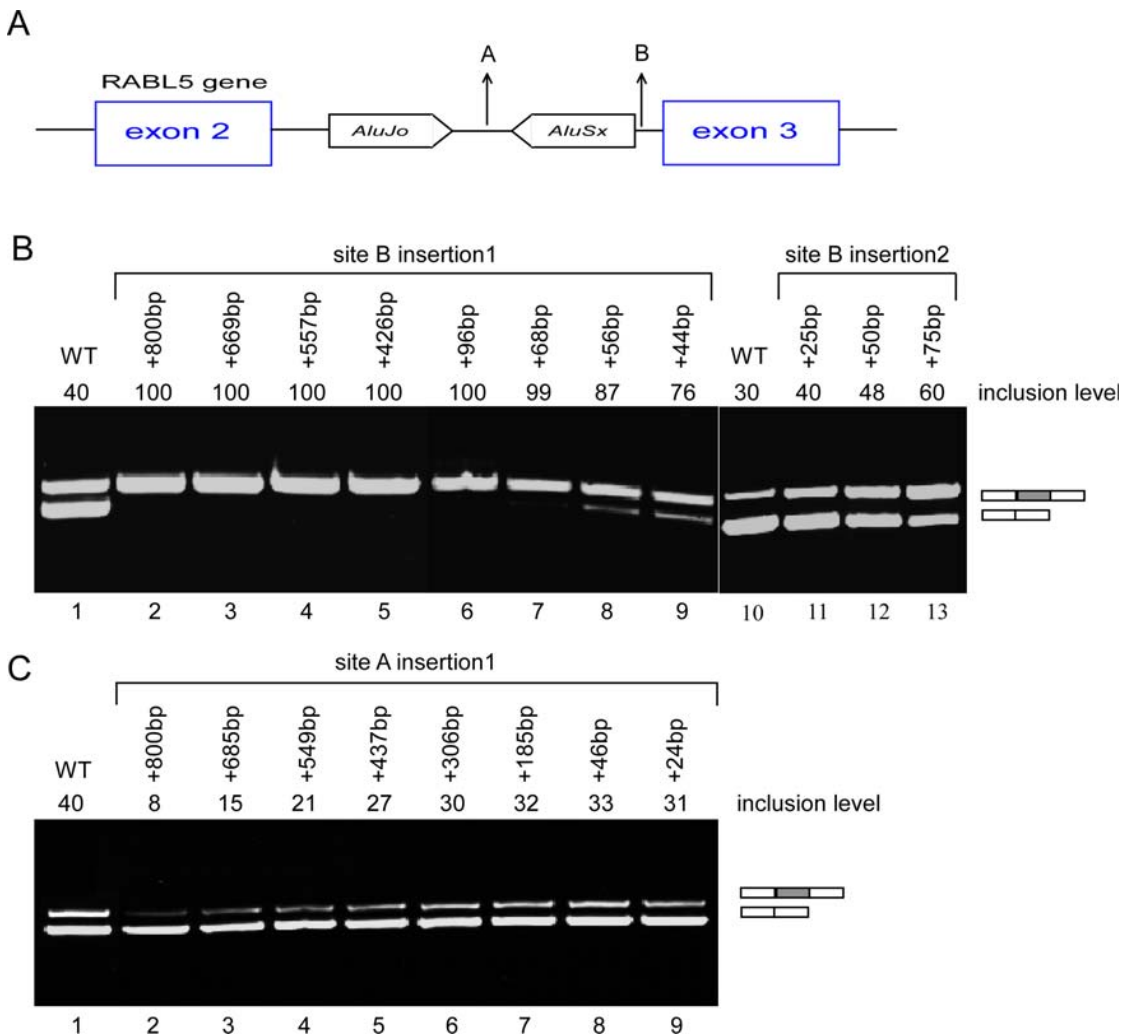

**Figure 4. Distance effect of *Alu* elements on the alternative splicing pattern.** (A) A schematic illustration of the genomic region between exons 2 and 3 of RABL5 gene. Arrows marked A and B indicate two positions where an intronic sequence was inserted. (B) An 800-nucleotide intronic sequence was inserted in site B. The 800-nucleotide sequence was gradually shortened to the size shown above each lane. The indicated wt and chimeric plasmids were transfected into human 293T cells, total RNA was collected and examined by RT-PCR analysis (lanes 1–9). Lanes 10–12 show insertions of a different sequence, containing 25 nucleotides without any known splicing regulatory sequences, into the same site. This sequence was duplicated and triplicated to generate 50- and 75-nucleotide inserts. These mutant RABL5 minigenes were examined as above. (C) Similar analysis as in panel B, except that the 800-nucleotide sequence and the shorter sequences were inserted into site A. Spliced products are shown on the right and each PCR product was confirmed by sequencing. Splicing products were separated on a 2% agarose gel. Numbers on top of the gel indicate percentage of exon inclusion as determined using ImageJ software.
doi:10.1371/journal.pgen.1000204.g004

### Alu2 Regulates Alternative Splicing of Exon 3

Figure 2 demonstrates that *Alu*2 suppresses the inclusion of exon 3. We therefore analyzed the sequence of *Alu*2 to determine what regions might be critical for this effect. Figure 5A shows the sequence of *Alu*2 and the mutations made. We deleted each of the two arms of *Alu*2 separately (Figure 5B, lanes 6 and 8) and mutated putative splicing signals (Figure 5B, lanes 2–5, 9–11). We found that the left arm of *Alu*2 is involved in the constitutive-to-alternative shift. Deletion of the left arm enhanced inclusion, whereas deletion of the right arm caused only a marginal effect (Figure 5B, compare lanes 1, 6, and 8). Mutations in the putative splicing signals in the right and left arms of *Alu*2 did not affect the splicing pattern (Figure 5B, lanes 2–5, 9–13). Our results do not support the possibility that the left arm functions as a pseudo-exon that abolishes or reduces selection of the exon 3 by competing with splicing factors (see also [58,61]). However, analysis of this data is not straight forward, because deletion of the entire right arm together with the left-arm-polypyrimidine tract (LPPT), which produced a short *Alu*2 sequence, caused complete skipping of exon 3 (Figure 5B, lane 7). Insertion of a complementary sequence to the short *Alu*2 sequence along with its upstream intronic sequence (to complete an *Alu*-like length of 280 nucleotides), 105bp upstream to the original short *Alu*2 (mimicking the original distance of *Alu*2 from *Alu*1), either in the sense or antisense orientation, did not affect full skipping of exon 3. Deletion of the right arm alone or the LPPT alone had a marginal effect on splicing of exon 3 (Figure 5B, compare lanes 6 and 9 to 7). These results imply that multiple sequences along *Alu*2 combine to suppress the recognition of exon 3.

Based on the location of the editing sites shown in figure 3B we concluded that in a large part of the left arm of *Alu*2 there is no editing, suggesting that this part might not participate in a dsRNA structure. Within the left arm we identified a potential sequence, which is not part of the *Alu*1-*Alu*2 putative pairing alignment. This region has the potential to form an internal stem-loop structure (Figure 5C, upper alignments). Deletion of this region, replacement of this sequence with a similar stem-loop structure of a different sequence, creation of fully paired stem structure, or disruption of the stem structure all caused full exon 3 skipping. These results imply that the sequence, rather than the potential secondary structure, of this region is important for the inclusion of exon 3. This sequence contains two putative SC35 binding sites.

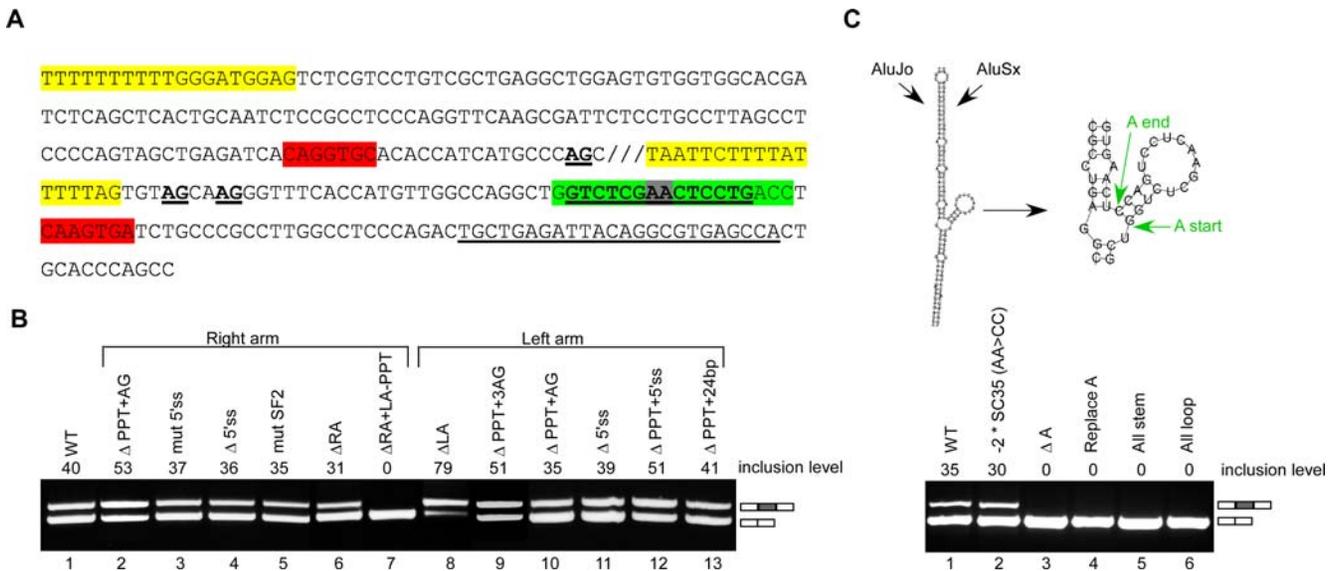

**Figure 5. The effect of *Alu*2 on the alternative exon.** (A) The sequence of the antisense *Alu*2. Mutated putative 5′ss is shown in red. A sequence of 24 nucleotides that was deleted is underlined. Three putative 3′ss that were mutated are in bold and underlined. In yellow are the right and left PPT regions with the downstream AG that were deleted. The green sequence is a stem and loop region of 18 nucleotides that was examined as shown in panel C (referred as 'A' region). Underlined in that region are two overlapping SC35 potential binding sites (the gray in the middle indicates the overlap region). (B) The indicated wt and mutant plasmids were transfected into 293T cells, total RNA was extracted, and splicing products were separated on a 2% agarose gel after RT-PCR analysis. Lane 1, splicing products of wt *RABL5* minigene. Lanes 2–13, splicing products of the indicated mutants. The two mRNA isoforms are shown on the right. Numbers on top of the gel indicate percentage of exon inclusion, as determined using ImageJ software. (C) The upper part illustrates the putative secondary structure formed by *Alu*Jo and *Alu*Sx, as predicted using the Vienna secondary structure web site (http://www.tbi.univie.ac.at/RNA). The green arrows in the right panel indicate the start and end positions of the stem and loop structure, marked as 'A'. The lower part shows the effect of the wt and mutant plasmids that were analyzed, as in panel B. Lane 1, wilt type. Lane 2, elimination of the two SC35 putative binding sites by mutating their overlapping sequence (AA were mutated to CC). Lane 3, deletion of the entire stem-loop 'A' sequence. Lane 4, replacement of the 'A' sequence by a random sequence lacking putative splicing binding sites. Lane 5, replacement of the loop sequence by a perfect complementary sequence, so that the 'A' element is entirely in a stem structure. Lane 6, disruption of the stem part of 'A', so that 'A' is entirely in a loop structure.
doi:10.1371/journal.pgen.1000204.g005

However, mutations that eliminated these potential binding sites without generating another known splicing regulatory sites had no effect on splicing of exon 3, indicating that this is not the sequence involved in the regulation (Figure 5C, lane 2). There are no other potential binding sites for known splicing regulatory factors in this sequence (based on [62–64]). Finally, addition of a potential complementary sequence to *Alu*1 did not effect splicing of exon 3 (not shown). Although *Alu*2 functions primarily to inhibit exon 3 selection, the above sequence within *Alu*2 enhances the inclusion of exon 3. Formation of a duplex between *Alu*1 and *Alu*2 is needed in order to present this intronic enhancer sequence properly for its effect on splicing of exon 3.

## Discussion

There are over 0.5 million copies of *Alu* elements in introns of human protein coding genes [65], yet their function in regulation of gene expression is largely unknown. Here we show that intronic *Alu*s are not 'neutral' elements; they affect splicing of flanking exons. Some of these effects can be directly linked to the shift from constitutive to alternative splicing during primate evolution. The regulation demonstrated here involves both positive and negative effects of *Alu* element in antisense orientation, in close proximity, and upstream to the regulated exon. This complex regulation causes the downstream exon to shift from constitutive to alternative splicing. There are several examples [53–57] of de novo insertions of *Alu* elements within introns that result in skipping of the adjacent exons. In three of the reported cases, the insertion of the *Alu* in the antisense orientation caused a total skipping of the adjacent exon.

Exon 3 of *RABL5* gene, analyzed in this study, is alternatively spliced in human and constitutively spliced in mouse, rat, dog, chicken, and zebrafish. Six *Alu*s have been inserted into the flanking introns of exon 3 since the last common ancestor of human and mouse. *Alu*2 was inserted in the antisense orientation just upstream of exon 3 and functions as a negative element that suppresses exon 3 selection. This negative effect is partially reversed by another *Alu* present in the same intron in the sense orientation. Although we were not able to fully resolve the mechanism by which the two *Alu*s regulate alternative splicing of the downstream exon, we provide evidence that regulation requires the formation of a double-stranded region between the two *Alu*s and a combination of negative and positive sequences located in *Alu*2. The end result of this complex regulation is a shift from constitutive to alternative splicing of the downstream exon. This results in a new primate-specific mRNA isoform that could acquire novel functions, as well as maintaining the original mRNA. Moreover, such a shift could introduce a premature termination codon resulting in truncated proteins that might have regulatory roles at certain times or in specific cell types as could be delicately determined by the levels of splicing regulatory proteins [59,60].

Introns in humans are considerably longer than their mouse counterparts, mostly due to the presence of *Alu* elements [66]. Introns that flank alternatively spliced exons are longer than introns that flank constitutively spliced ones [4,67]. We found a correlation between the splicing pattern of exons and the presence of *Alu*s in the flanking introns. First, alternatively spliced exons are flanked by introns containing more *Alu*s compared with introns flanking constitutively spliced ones, even when controlling for the difference in intron lengths. Second, more *Alu*s are present in human introns than are corresponding mouse B1 elements in the orthologous mouse introns. This second observation correlates with the finding that there are more species-specifically, alterna- tively spliced exons in human than in mouse (354/612 alternatively spliced events in human and 258/612 alternatively spliced events in mouse; $\chi^2$, p –value <0.01).

It was suggested that intron complementarities formed by multiple copies of *Alu* could help define and increase the splicing efficiency of very large metazoan introns [68]. However, it may be also possible that formation of a long and stable double-stranded structure in the upstream intron, especially near the splice site as in the case studied in this manuscript, reduces the ability of the splicing machinery to properly recognize the downstream exon, leading to slower splicing kinetics or suboptimal exon selection and, thus, to intron retention or exon skipping. Supporting the hypothesis that formation of dsRNA in introns might slow splicing is a recent publication showing that formation of dsRNA during pre-microRNA formation can slow splicing of the intron where the microRNA resides [69].

Although the effect of intronic retroelements on the splicing of flanking exons is presumably not a general trend that applies to all exons, it is relevant to a certain fraction of alternatively spliced exons (1.5% to 4% of the alternatively skipped exons in human).

Our analysis indicated that the presence of *Alu* elements is correlated with the mode of splicing of adjacent exons. There is an 'exclusion zone' in intron sequences flanking exons, where insertion of *Alu* elements is presumably under purifying selection. The length of this 'exclusion zone' is similar to that of the human-mouse conserved sequences flanking alternatively spliced exons (~80–150 nucleotides). This is presumably indicative of regions where the presence of intronic splicing regulatory sequences can affect alternative splicing of the adjacent exon [5,18,70,71]. *Alu*s might be excluded from the proximal intronic sequences flanking constitutively spliced exons because *Alu*s were never inserted into these regions or because *Alu*s were inserted in an equal proportion in all gene regions (intronic and exonic) but we currently observe only those *Alu*s that have escaped purifying selection. The major burst of *Alu* retroposition took place 50–60 million years ago and has since dropped to a frequency of one new retroposition for every 20–125 new births [72,73]. As some of these insertions were deleterious and thus selected against, we probably detect intronic *Alu*s that are neutral, mildly deleterious, or beneficial to human fitness. Some of these beneficial intronic *Alu*s presumably altered splicing of the flanking exons and resulted in the generation of new isoforms that presented an advantage during primate evolution and were thus fixated in our genome. The research described here sheds light on how *Alu* elements have shaped the human genome.

## Materials and Methods

### Dataset Compilation

A dataset of 596 alternatively spliced exons, conserved between human and mouse, was derived from a previously compiled dataset [5]. In addition, 45,553 human-mouse conserved constitutively spliced exons were obtained from Carmel et al. [74]. Species-specific exons (354) were extracted from a dataset of 4,262 human-mouse orthologous exons that are suspected to splice differently in human and mouse based on initial EST analysis [74]. For details of how the datasets were built see [17].

### Human and Mouse Intronic Dataset

Introns and exons for human (*Homo sapiens*, Build 35.4) and mouse (*Mus musculus*, Build 34.1) were extracted from the Exon-Intron Database (http://hsc.utoledo.edu/depts/bioinfo/database.html) [75]. These intron sequences were analyzed with Repeat-Masker software version 3.1.0 [76] (www.repatmasker.org) using Repbase update files [77].

### Retrotransposed Elements Analysis

Since Alus are primate specific, the distribution was computed only from human flanking introns. The density of Alu elements was calculated per 1000-bp intron length according to the following equation:

$$Alu_{density} = \frac{N \times 1000}{L}$$

N = number of Alus within the intron; L = the length of the intron; $Alu_{density}$ = Alu density. For the detection of retrotransposed elements, we used the RepeatMasker (http://www.repeatmasker.org) software version 3.1.0 [76] and Repbase update [77].

### Statistical Analysis

A T-test was used to calculate statistical differences between two populations; for $\chi^2$ test with 2×2 contingency table, Fisher's exact test was used.

### The Potential Alternatively Skipped Exons under Intronic Alu Regulation

The alternatively skipped exons in the human genome (build hg18) were extracted by downloading the knownAlt table from UCSC genome browser [48]. The presence or absence of Alu within the upstream intron was determined using RepeatMasker tables downloaded from UCSC. The conservation of these introns was analyzed using MAF pairwise alignments between the human genome (build hg18) and the mouse genome (build mm9) downloaded from UCSC genome browser. The intersections between these tables were done using the Galaxy sever [49].

### Bioinformatic Analysis of Alu Editing within Intronic Sequences

We extracted data of on 10,113 nucleotides that undergo mRNA editing in the human genome from Levanon et al. [24]. The UCSC genome browser [58] was then used to extract data of all Alu elements in the human genome (build hg18) using the RepeatMasker [76] annotations, and to extract human RefSeq intronic sequences. Intersection of these three datasets yielded 953 Alu elements that are embedded in intronic sequences and undergo mRNA editing.

### Plasmid Constructs

The RABL5 (RAB member RAS oncogene family-like 5) minigene was generated by amplifying a human genomic fragment using PCR reaction. Each primer contained an additional sequence encoding a restriction enzyme. The PCR product was restriction digested and inserted into the pEGFP-C3 plasmid (Clontech) and sequenced to confirm that the desired construct was obtained. The RABL5 minigene, contains exons 2 through 4 (2.7 kb). The intron replacements with the RABL5 Alu1, Alu2, and Alu3 were done by PCR opening of the plasmid lacking the specific Alu (#1, #2 or #3) and ligation with a fragment of 270 intronic-nucleotides taken from a PCR amplification directed to the IKBKAP gene intron number 20 (primer forward, 5′AGAATCGT-GACACTCATCATATAAAGGAGG3′; and primer reverse, 5′CAAAACATTAGTATAGATCTTTCCAATACA3′). The 800-nucleotide insertion1 was taken from PCR amplification directed to the IMP gene intron number 11 (primer forward, 5′ATCACTCTGCACTTTCTCCCAT3′; primer reverse 5′AC-CATGTCCACTTCATCCAGTTC3′). Insertion2 is a 25-bp sequence, free of any known splicing regulatory elements, that was doubled or tripled into 50-bp and 75-bp sequences, (5′CTATCTGATAAGCTGCGAGCAATT3′).

### cDNA Amplification

Endogenous PCR amplification was done on a cDNA template originating from a neuroblastoma cell-line (SH-SY5Y). Amplification was performed for 30 cycles, consisting of denaturation for 30 seconds at 94°C, annealing for 45 seconds at 52°C or 56°C, and elongation for 1 minute at 72°C. The products were separated in a 1.5% agarose gel. The upper PCR product was Topo-ligated (Invitrogen) and sequenced. The primers used were: forward (exon 2), 5′CAGAATCTTCTGACATCACTG3′; or forward (intron 2), 5′GTGAGCCCTGACAAATCTGTGT3′; and reverse (exon 3) 5′GTTGCTGGTAACATGCGGGTTC3′.

### Site-Directed Mutagenesis, Transfection, RNA Isolation, and RT-PCR Amplification

For details see [78].

### Supporting Information

**Figure S1** Intron length distribution of human introns.
Found at: doi:10.1371/journal.pgen.1000204.s001 (0.17 MB DOC)

**Figure S2** A screen shot created by the UCSC genome browser.
Found at: doi:10.1371/journal.pgen.1000204.s002 (1.44 MB DOC)

**Figure S3** Potential dsRNA of AluJo+ and AluSx-.
Found at: doi:10.1371/journal.pgen.1000204.s003 (0.51 MB DOC)

**Text S1** Minigenes' sequence.
Found at: doi:10.1371/journal.pgen.1000204.s004 (0.03 MB DOC)

**Text S2** 270 nucleotides non-Alu intronic sequence from intron 20 of IKBKAP gene.
Found at: doi:10.1371/journal.pgen.1000204.s005 (0.02 MB DOC)

**Text S3** 800 nucleotides intronic sequence from intron 11 of IMP gene.
Found at: doi:10.1371/journal.pgen.1000204.s006 (0.02 MB DOC)

### Author Contributions

Conceived and designed the experiments: GLM OR GA. Performed the experiments: GLM OR. Analyzed the data: GLM OR EK NS AG GA. Contributed reagents/materials/analysis tools: GLM GA. Wrote the paper: GLM EK NS EYL GA.